\documentclass[review]{elsarticle}

\usepackage{lineno,hyperref}

\usepackage{amsmath}
\usepackage{amsbsy}
\usepackage{amssymb}
\usepackage{color}
\usepackage{epsfig}
\usepackage{setspace}
\usepackage{bm}

\newcommand{\prob}{P}
\newcommand{\mbf}{\mathbf}
\newcommand{\bs}{\boldsymbol}
\newcommand{\E}{I\!\!E\hspace{.03cm}}

\modulolinenumbers[5]

\journal{ArXiv}

\begin{document}

\begin{frontmatter}

\title{Flexible multi-state models for  interval-censored data: specification, estimation, and an application to ageing research}

\author{Robson J. M. Machado} 
\ead{robson.machado.14@ucl.ac.uk}
\author{Ardo van den Hout }
\address{Department of Statistical Science, University College London, 1-19 Torrington Place, \\ London WC1E 7HB, UK.} 





\begin{abstract}
Continuous-time multi-state survival models can be used to describe health-related processes over time. In the presence of interval-censored times for transitions between the living states, the likelihood is constructed using transition probabilities. Models can be specified using parametric or semi-parametric shapes for the hazards. Semi-parametric hazards can be fitted using $P$-splines and penalised maximum likelihood estimation. This paper presents a method to estimate flexible multi-state models which allows for parametric and semi-parametric hazard specifications. The estimation is based on a scoring algorithm. The method is illustrated with data from the English Longitudinal Study of Ageing.\end{abstract}


\end{frontmatter}


\section{Introduction}
\label{S:1}

Multi-state models are routinely used in research where change of status over time is of interest. In epidemiology and medical statistics, the models are used to describe health-related processes over time, where status is defined by a disease or a condition. In social statistics and in demography, the models are used to study processes such as region of residence, work history, or marital status. A multi-state model which includes a dead state is called a multi-state survival model.

The specification of a multi-state survival model depends partly on the study design which generated the longitudinal data that are under investigation. An important distinction is whether or not exact times are observed for transitions between the states. This paper considers study designs where death times are known exactly (or right censored) and where transition times between the living states are interval censored. Many applications in epidemiology and medical statistics have this property as it is often hard to measure the exact time of onset of a disease or condition. Examples are dementia, cognitive decline, disability in old age, and infectious diseases.

A multi-state survival model describes change in a discrete longitudinal outcome variable {\it and} attrition due to death. If a longitudinal outcome variable can be adequately described by a set of states, then a multi-state survival model is an alternative to so-called joint models. An example of the latter is the shared-parameter joint model which consists of combining a survival model for the event time with a mixed-effects model for the longitudinal outcome  \cite{rizopoulos2012}.

This paper defines continuous-time multi-state survival models by specifying transition-specific hazard models. Time-dependency of the process is defined by using parametric and semi-parametric formulations in the specification of the baseline hazard functions. The semi-parametric specification is made with $P$-splines \cite{eilers1996}, which are $B$-splines with penalties on the difference of adjacents splines \cite{eilers1996}. Using $B$-splines is a general method for smoothing \cite{eilers1996, de1978practical}. 

Because the definition of the functional form of the hazard can be transition specific, model specification can cover a wide range of multi-state survival processes. The methodology presented in this research is for multi-state models with at least one transition hazard specified with $P$-splines. Estimation is carried out with penalised maximum likelihood, where the maximisation is undertaken by using a Fisher scoring algorithm. This algorithm is an extension of the work by Jennrich and Bright \cite{jennrich1976} and Kalbfleisch and Lawless \cite{kalbfleisch1985}.

The models are formulated in a Markov process framework. Time-dependency is approached by using a piecewise-constant approximation and defining a series of time-homogeneous processes. For each of these homogeneous processes, the solution to the Kolmogorov forward equations (a first-order differential equation) is computed using eigenvalue decomposition. The method can be applied to multi-state models with any number of states and specification of hazard transitions can vary across transitions. 


Semi-parametric multi-state models for interval-censored data have been discussed in the literature. Titman \cite{titman2011} uses a numerical approximation to calculate the transition probabilities. The advantage is that there is no need to define a grid for a piecewise-constant approximation, but it is computationally more demanding than using eigenvalue decomposition\textemdash especially in the case of continuous-scale covariates. Also, even though $B$-splines are used to model transitions intensities, the log-likelihood is maximised without penalisation. Joly and Commenges \cite{joly1999penalized} use a penalised approach for a progressive three-state model. Estimation is performed with an algorithm which uses derivatives of the penalised log-likelihood. The smoothing parameters of the model are selected using a grid search with cross-validation. Joly \textit{et al.} \cite{joly2002penalized} use the same approach for an illness-death model. The method used in both papers requires explicit expressions for the transition probabilities. Calculating these formulas can be intractable for more complex models, such as models with more than four states and recovery \cite{jackson2011multi}.

Sennhenn-Reulen and Kneib \cite{sennhenn2016structured} developed an estimation procedure for multi-state models based on a structured lasso penalisation. The aim of their research is to identify covariate effects coefficient equal to zero. Baseline transition intensities are specified with piecewise-constant models or unspecified and equal across all transitions. Their method is not defined for interval-censored data. Therefore, their work is different from ours in scope and methodology. 

\begin{figure}
 \centerline{\includegraphics[width=11cm]{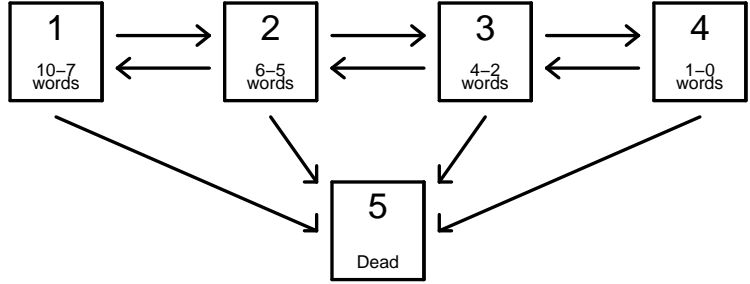}}
\caption{Five-state model for longitudinal data in ELSA on number of words remembered in a recall \label{figAvdH:ELSAdiagram}
}
\end{figure}

To illustrate the statistical modelling and the penalised maximum likelihood estimation, longitudinal data on survival and change of cognitive function in older population is analysed. The data stem from the English Longitudinal Study of Ageing (ELSA, {\tt www.elsa-project.ac.uk}) and the longitudinal response variable is the number of words remembered in a recall from a list of ten. Of interest is the effect of age and gender on cognitive change over time when controlling for education. Four states are defined by the number of words an individual can remember, see Figure \ref{figAvdH:ELSAdiagram}. The dead state is the fifth state. 
The transition times between the living states are interval-censored, but death times are known. We acknowledge that these data are also analysed in Van den Hout \cite{van2014multi}. This paper extends that analysis by using $P$-splines for hazard specification.

\section{Models}\label{sec:models}
For a continuous-time Markov chain  $Y(t)$ on finite state space $\mathcal{S}$, time-homogeneous {\it transition probabilities}
are given by \[
p_{rs}(t)=P\big(Y(t)=s|Y(0)=r\big),
\]
for $r,s \in \mathcal{S}$ and $t\geq 0$. Transition matrix  $\mathbf{P}(t)$ contains these probabilities such that the rows sum up to 1.  The Chapman-Kolmogorov equation is $\mathbf{P}(u+t)=\mathbf{P}(u)\mathbf{P}(t)$. The {\it transition intensities} (or {\it hazards}) are given by
\[
q_{rs}=\lim_{\Delta\downarrow 0}\frac{P\big(Y(t+\Delta)=s|Y(t)=r\big)}{\Delta},
\]
for $r\neq s$. The matrix with off-diagonal entries $q_{rs}$ and diagonal entries $q_{rr}= -\sum _{r\neq s} q_{rs}$ is the {\it generator matrix}  $\mathbf{Q}$. Given $\mathbf{Q}$, the solution for $\mathbf{P}(t)$ subject to $\mathbf{P}(0)= \mathbf{I}$ is $\mathbf{P}(t)=\exp(t \mathbf{Q} )$, see, e.g., Norris \cite{norris1998markov}. In general, the computation of the exponential of a square matrix is not straightforward, see Moler and Van Loan \cite{moler2003nineteen} for a discussion of methods and efficiency.

A time-dependent hazard regression model for transition intensities combines baseline hazards with log-linear regression and is given by
\begin{eqnarray}
q_{rs}(t)=q_{rs.0}(t)\exp\big(\boldsymbol{\beta}_{rs}^{\top}\mathbf{x}\big),
\label{eq:hazardmodel}
\end{eqnarray}
where $\mathbf{x}$ is a covariate vector without an intercept.  Transition-specific time dependency can be introduced via baseline hazards. Parametric examples are
\begin{eqnarray}
\mbox{Weibull:\ \ \ \ } q_{rs.0}(t)  &=& \beta_{rs.0}\tau_{rs} t^{\tau_{rs}-1} \hspace{1.2cm} \beta_{rs.0},\tau_{rs}>0 \\
\mbox{Gompertz:\ \ \ \ } q_{rs.0}(t) &=& \beta_{rs.0}\exp(\xi_{rs} t) \hspace{1cm} \beta_{rs.0}>0\ .
\end{eqnarray}
Semi-parametric models can be defined in a similar way. An example is using $P$-splines to allow for flexible modelling of the time-dependency. Applications to multi-state models can be found in Kneib and Hennerfeind \cite{kneib2008bayesian}. The  semi-parametric formulation with $P$-splines of the baseline hazard is
\[
q_{rs.0}(t)= \exp\left(\sum_{k=1}^{K}\alpha_{rs.k}B_k\right),
\]
where\textemdash in this case\textemdash the choice of the number of knots $K$ is the same for all transition, but the $\alpha$s are not. Flexible multi-state models can be defined by $P$-splines or a combinations of the hazard specifications above. 

\section{Penalised maximum likelihood estimation}
\subsection{Likelihood function}\label{sec:likelihood}
Given a multi-state survival model, maximum likelihood inference can be used to analyse longitudinal data. In the presence of interval censoring, the likelihood function is constructed using transition probabilities.
Let the state space be $\mathcal{S}=\{1,2,..,D\}$, with $D$ the dead state.

Consider a series of states $Y_1,...,Y_{n}$ observed at times $t_1,...,t_n$, respectively. The inference is conditional on the first observed state. For $Y_2,...,Y_{n}$, the distribution is
\begin{eqnarray}
\prob\left(Y_{n}=y_{n},...,Y_{2}=y_{2}|Y_{1}=y_{1}, \bs{\theta},  \mbf{t}, \mbf{X}\right),\label{eq:jointdistr}
\end{eqnarray}
where ${\bs{\theta}}$ is the vector with the model parameters, $\mbf{t}=(t_1,...,t_n)^\top$, and the $n\times p$ matrix $\mbf{X}$ contains the values of the $p$ covariates at each of the $n$ time points. A conditional first-order Markov assumption is used to define the distribution (\ref{eq:jointdistr}) of $Y_2,...,Y_{n}$ as
\begin{eqnarray*}
\prod_{j=2}^{n}\prob\left(Y_j=y_j|Y_{j-1}=y_{j-1},\bs{\theta}, t_{j-1}, \mbf{x}_{j-1}\right),
\end{eqnarray*}
where $ \mbf{x}_{j-1}$ is the $(j-1)^{th}$ row in $\mbf{X}$.

Next consider an individual $i$ with observed values $y_1,...,y_{n-1} \in\mathcal{S}\backslash D$, and a last observation $y_{n}$ which is either a value in $\mathcal{S}$ or a code for right-censoring.
The likelihood contribution for this individual is $L_i=\prod_{j=2}^n L_{ij}$, where
\begin{eqnarray}
L_{ij}=
\left\{
\begin{array}{lll}
P\left(Y_j=y_j|Y_{j-1}=y_{j-1},{\bs{\theta}}, t_{j-1}, \mbf{x}_{j-1}\right) & \mbox{for} & j=2, ..., n-1\\
C(y_{n}|y_{{n}-1})  & \mbox{for} & j=n\ .
\end{array}
\label{eqAvdH:lik}
\right.
\end{eqnarray}
If a living state at $t_{n}$ is observed, then
$C(y_{n}|y_{n-1})=P\left(Y_{n}=y_{n}|Y_{n-1}=y_{n-1}\right)$, where part of the conditioning is ignored in the notation.
If the state is right censored at $t_{n}$, then $C(y_{n}|y_{n-1})=\sum_{s=1}^{D-1}P\left(Y_{n}=s|Y_{n-1}=y_{n-1}\right)$.
If the state at $t_{n}$ is $D$, then known time of death is taken into account by defining
\[
C(y_{n}|y_{n-1})=\sum_{s=1}^{D-1}P\left(Y_{n}=s|Y_{n-1}=y_{n-1}\right)q_{sD}(t_{n-1}).
\]
Given $N$ individuals, the log-likelihood function is given by
\begin{eqnarray}
\ell(\bs{\theta})=\sum_{i=1}^{N}\ \log L_{i}=\sum_{i=1}^{N}\sum_{j=2}^{n_i}\ \log L_{ij},
\label{eq:likelihood}
\end{eqnarray}
where $n_i$ is the number of observation times for individual $i$.

Above definition of the likelihood function can also be found in Jackson \cite{jackson2011multi}. Including time-dependency as defined by the models in Section \ref{sec:models}, does not affect the basic structure of the likelihood function. Similar expressions of the likelihood function can be found in Kalbfleisch and Lawless \cite{kalbfleisch1985}, Kay \cite{kay1986}, and Gentleman \textit{et al.} \cite{gentleman1994}.

\subsection{Penalised log-likelihood function}
For the semi-parametric multi-state model, at least one baseline hazard function is specified with $P$-splines. If a transition is defined by $P$-splines, we specify a large set of equidistant knots. To control the smoothness of the estimated curve, a penalty based on finite differences of the coefficient of adjacent $P$-splines is imposed to the log-likelihood function. Without loss of generality, suppose there are $K$ knots for each smoothed hazard. Let $\mbox{\boldmath$\beta$}$, $\mbox{\boldmath$\alpha$}$ and $\mbox{\boldmath$\xi$}$ represent the vector of parameters associated to the parametric, semi-parametric and covariates components of a multi-state model, respectively. Let $\mbox{\boldmath$\theta$}^\top = (\mbox{\boldmath$\beta$}^\top, \mbox{\boldmath$\alpha$}^\top, \mbox{\boldmath$\xi$}^\top)$ be the full set of parameters and $l(\mbox{\boldmath$\theta$})$ be the log-likelihood of a semi-parametric multi-state model. The penalised log-likelihood function is 
\begin{eqnarray}
\ell_p(\mbox{\boldmath$\theta$}) &=&  \ell(\mbox{\boldmath$\theta$}) -  \frac{1}{2} \sum _{j=1}^{s}\lambda_j \mbox{\boldmath$\alpha$}_j^\top \mathbf{D}_j^\top \mathbf{D}_j \mbox{\boldmath$\alpha$}_j \nonumber \\
&=& \ell(\mbox{\boldmath$\theta$}) - \frac{1}{2} \mbox{\boldmath$\theta$}^\top \mathbf{J}(\mbox{\boldmath$\lambda$}) \mbox{\boldmath$\theta$}, \label{eq:penlike}
\end{eqnarray}
where $\mbox{\boldmath$\alpha$}_j = (\alpha_{j1}, \ldots, \alpha_{jK})^\top$, $\mbox{\boldmath$\lambda$}$ is the vector of smoothing parameters, $\mathbf{D}$ is the matrix representation of the difference operator $\Delta$ of adjacent $P$-splines \cite{eilers1996} and $ \mathbf{J}(\mbox{\boldmath$\lambda$})$ is the penalty matrix. $\mathbf{J}(\mbox{\boldmath$\lambda$})$ is a block diagonal matrix with blocks $\lambda_j \mathbf{D}^\top \mathbf{D}$ for penalising $P$-splines parameters and zeros elsewhere \cite{gray1992flexible}.

\subsection{Piecewise-constant hazards}
Let $\mathbf{P}(t,t +\Delta)$ denote the transition matrix for any time interval $(t,t+\Delta]$. Time-dependency in  hazard model (\ref{eq:hazardmodel}), implies that $\mathbf{P}(t_1,t_1+\Delta)\neq \mathbf{P}(t_2,t_2+\Delta)$ for $t_1\neq t_2$.

Time-dependency of the hazard can be taken into account by using a piecewise-constant approximation.
Given consecutive times $t_1, t_2, \dots t_n$, define the transition matrix for $(t_1,t_n]$  by 
\[
\mathbf{P}(t_1,t_n)=\mathbf{P}(t_1,t_2)\times \cdots\times \mathbf{P}(t_{n-1},t_n),
\]
where the matrices at the right-hand side are derived using generator matrices $\mathbf{Q}(t_1), \mathbf{Q}(t_2), \dots,\mathbf{Q}(t_{n-1})$, respectively.

In longitudinal data for continuous-time models, follow-up times often vary across individuals. If that is the case, the individual-specific follow-up times can be used to define the piecewise-constant approximation for the individual likelihood contributions. This implies that a transition probability such $P\left(Y_{j}=y_{j}|Y_{j-1}=y_{j-1}\right)$ is derived by using $\mathbf{Q}(t_{j-1})$ to compute $\mathbf{P}(t_{j-1},t_j)$.

Instead of letting the data determine the grid for the piecewise-constant approximation, it is also possible to impose a  grid,  which is the same for all individual likelihood contributions \cite{van2008}. In this case,
time intervals in the data are embedded in the grid. For example, say the grid is defined by $u_1,...,u_M$. For an observed time interval $(t_1,t_2]$, determine $j_1$ and $j_2$ such that $u_{j_1}<t_1\leq u_{j_1+1}$ and $u_{j_2}<t_2\leq u_{j_2+1}$. The transition matrix for $(t_1,t_2]$ is then defined by
\[
\mbf{P}(t_1,t_2)=\mbf{P}\left(t_1,u_{j_1+1}\right)\mbf{P}\left(u_{j_1+1},u_{j_1+2}\right)\times  \cdot\cdot\cdot \times \mbf{P}\left(u_{j_2},t_2\right),
\]
using generator matrices $\mbf{Q}(u_{j_1}), \mbf{Q}(u_{j_1+1}), \dots, \mbf{Q}(u_{j_2})$, respectively. For this approach covariate values are needed at all grid points $u_1,...,u_M$. For a covariate with a stochastic time-dependency, these values may not be available in the data.

\subsection{Scoring algorithm}
Given a piecewise-constant approximation to the time-dependency in the hazard model (\ref{eq:hazardmodel}), a scoring algorithm can be used to maximise  the logarithm of the likelihood function (\ref{eq:penlike}).
A scoring algorithm solves maximum likelihood equations numerically by iteratively estimating a root of the first-order derivative of the log-likelihood function. The first-order derivative of the log-likelihood function is called the {\it score function}.

As in Section \ref{sec:likelihood}, let ${\bs{\theta}}=(\theta_1,...,\theta_q)$ be the vector with model parameters. The
crucial step is to derive $\partial\mathbf{P}(t_1,t_2)/\partial \theta_k$ for a given time interval $(t_1,t_2]$. The important aspects of the scoring algorithm are
\begin{itemize}
\item[{\it (i)}] Because of the piecewise-constant approximation, the basic formulas for the time-homogeneous case in Kalbfleisch and Lawless \cite{kalbfleisch1985} apply to the constituent intervals with constant hazards in the likelihood function.
\item[{\it (ii)}] By using an eigenvalue decomposition of a generator matrix $\mathbf{Q}(t)$, only the derivatives $\partial\mathbf{Q}(t)/\partial \theta_k$ are needed \cite{jennrich1976}.
\item[{\it (iii)}] For the Weibull and the Gompertz hazard models, and for a model with $P$-splines, derivatives $\partial\mathbf{Q}(t)/\partial \theta_k$ are straightforward to derive.
\item[{\it (iv)}] The likelihood contributions for exact death times and right-censoring are made up of transition probabilities and transition hazards and can be dealt with by using {\it (i)} -- {\it (iii)}.

\end{itemize}

To specify the scoring algorithm, the derivative of a transition matrix is presented first. Given piecewise-constant intensities, the likelihood contribution for an observed time interval $(t_1,t_2]$ is defined using a constant generator matrix $\mbf{Q}=\mbf{Q}(t_1)$. For the eigenvalues of $\mbf{Q}$ given by
$\mbf{b}=(b_1,...,b_D)$, define $\mbf{B}=\mbox{diag}(\mbf{b})$. Given matrix $\mbf{A}$ with the eigenvectors as columns, the eigenvalue decomposition is $\mbf{Q}=\mbf{A}\mbf{B}\mbf{A}^{-1}$.  The transition probability matrix $\mbf{P}(t)=\mbf{P}(t_1,t_2)$ for elapsed time $t=t_2-t_1$ is given by
\[
\mbf{P}(t)=\mbf{A}\ \mbox{diag}\left(e^{b_1t},...,e^{b_Dt}\right)\ \mbf{A}^{-1}.
\]

As described in Kalbfleisch and Lawless \cite{kalbfleisch1985}, the derivative of $\mbf{P}(t)$ can be obtained as
\[
\frac{\partial}{\partial \theta_k}\mbf{P}(t)=\mbf{A}
\mbf{V}_k
\mbf{A}^{-1},
\]
where $\mbf{V}_k$ is the $D \times D $ matrix with $(l,m)$ entry
\begin{eqnarray*}
\left\{
\begin{array}{ll}
 g_{lm}^{(k)}\left[\exp(b_lt)-\exp(b_mt)\right]/(b_l-b_m) & l\neq m\\\\
 g_{ll}^{(k)}t\exp(b_lt)  & l= m,
\end{array}
\right.
\end{eqnarray*}
where $g_{lm}^{(k)}$ is the $(l,m)$ entry in $\mbf{G}^{(k)}=\mbf{A}\partial \mbf{Q}/\partial \theta_k\mbf{A}^{-1}$.

For the parametric and semi-parametric time-dependent hazard models in Section \ref{sec:models}, matrix $\partial \mbf{Q}(t_1)/\partial \theta_k$ is straightforward to derive.

The scoring algorithm can now be defined as follows. Let the $q\times 1$ vector $\mbf{S}(\bs{\theta})$ denote the score function.
The $k$th entry of $\mbf{S}(\bs{\theta})$ is given by
\[
\sum_{i=1}^N\sum_{j=2}^{n_i}\frac{\partial}{\partial \theta_k}\log L_{ij}\,.
\]
The expected observed information matrix is called the Fisher information and is given by
$\bs{\mathcal{I}}(\bs{\theta})=\E\left[\mbf{S}(\bs{\theta})\mbf{S}(\bs{\theta})^\top \right]$, which can be estimated by
defining the $q\times q$ matrix $\mbf{M}({\bs{\theta}})$ with $(k,l)$ entry
\[
\sum_{i=1}^N\sum_{j=2}^{n_i}\frac{\partial}{\partial \theta_k}\log L_{ij}
\frac{\partial}{\partial \theta_l}\log L_{ij}\,.
\]

The penalised score $\mbf{S}_{p}(\bs{\theta})$ and estimated penalised Fisher information matrix $\mbf{M}_{p}(\bs{\theta})$ are given by 
\begin{eqnarray*}
\mbf{S}_{p}(\bs{\theta}) &=& \mbf{S}(\bs{\theta}) - \mathbf{J}(\bs{\lambda})\bs{\theta},\\
\vspace{4cm}
\mbf{M}_{p}(\bs{\theta}) &=& \mbf{M}(\bs{\theta}) + \mathbf{J}(\bs{\lambda}).
\end{eqnarray*}

Given starting values $\bs{\theta}^{(0)}$, the scoring algorithm is given for $v=1,2,3\dots$ by
\begin{eqnarray*}
\bs{\theta}^{(v+1)}= \bs{\theta}^{(v)}+\mbf{M}_p\big(\bs{\theta}^{(v)}\big)^{-1}\mbf{S}_p\big(\bs{\theta}^{(v)}\big)\,.
\end{eqnarray*}

Let $\bs{\mathcal{I}}_p(\bs{\theta})$ represents the penalised Fisher information matrix. The asymptotic covariance matrix of the penalised maximum likelihood estimate $\widehat{\bs{\theta}}$ is equal to $\bs{\mathcal{I}}_p(\bs{\theta})^{-1}$. Hence, after convergence, the covariance matrix of the penalised maximum likelihood estimate $\widehat{\bs{\theta}}$ is estimated by $\mbf{M}_p(\widehat{\bs{\theta}})^{-1}$ \cite{radice2016copula}.


\section{Estimation of smoothing parameter}

Estimating optimal value for the smoothing parameters $\mbox{\boldmath$\lambda$}$ is crucial for fitting models with splines   
\cite{gu1991minimizing}. A common method for choosing smoothing parameters is the Akaike Information Criterion (AIC). The AIC definition is equivalent to  
\begin{eqnarray*}
\mbox{AIC}(\mbox{\boldmath$\lambda$}) = -2\ell_p + 2df.
\end{eqnarray*}
The degrees of freedom $df$ is a measure of model complexity. For parametric models, the degrees of freedom are equal to the number of independent parameters in the model. For semi-parametric models with $P$-splines, these can be defined as 
\begin{eqnarray*}
df(\mbox{\boldmath$\lambda$}) = \mbox{tr}[\mathbf{M}(\mathbf{M} + \mathbf{J}(\mbox{\boldmath$\lambda$}))^{-1}],
\end{eqnarray*}
where $\mathbf{M}$ is the (estimated) Fisher information matrix and $\mathbf{J}$ is the penalty function \cite{gray1992flexible}. Small values of $\mbox{\boldmath$\lambda$}$ lead to wiggly functions, while large values lead to more conservative estimated functions that tend to a straight line.

\section{Prediction}
Once a multi-state model is fitted using a parametric and semi-parametric hazard model, estimated model parameters can be used for prediction. Typically this concerns computing transition matrices as a function of the penalised maximum likelihood estimate.
The covariance of a function of model parameters can be estimated by Monte Carlo simulation or by using the multivariate delta method, see also Titman \cite{titman2011}. Because transition probabilities are restricted to $[0, 1]$, using simulation is recommended as the default method. The delta method does not take the restriction into account and this can have a substantial knock-on effect on long-term prediction. This paper focuses on the simulation method. 

Let $\widehat{\mathbf{V}}_{\boldsymbol{\theta}}$ denote the estimated covariance matrix of the penalised maximum likelihood estimate $\widehat{\boldsymbol{\theta}}$. Of interest is the estimation of $\mathbf{P}(t_1,t_2)$ for arbitrary $t_1$ and $t_2>t_1$.
In the case of a time-dependent model, let the grid for the piecewise-constant approximation be defined by $u_{j+1}=u_j+h$ for $j=1,...,M$ such that $u_1=t_1$ and $u_M=t_2$. Given this grid, matrix $\mathbf{P}(t_1,t_2)$ is estimated by $\mathbf{P}(u_1,u_2) \times \cdot\cdot\cdot \times  \mathbf{P}(u_{M-1}, u_M)$.

For Monte Carlo simulation, parameter vectors ${\boldsymbol{\theta}}^{(b)}$ are drawn from $N(\widehat{\boldsymbol{\theta}},\widehat{\mathbf{V}}_{\boldsymbol{\theta}})$, for $b=1,...,B$, and for each sampled  ${\boldsymbol{\theta}}^{(b)}$, $\mathbf{P}(t_1,t_2)$ is calculated. Summary statistics such
as mean and covariance can be derived easily from the $B$ realisations of $\mathbf{P}(t_1,t_2)$.

Sampling from a $K$-variate normal distribution $N(\bs{\mu},\bs{\Sigma})$ is possible by using the Cholesky decomposition $\bs{\Sigma}=\mbf{L}\mbf{L}^{\top}$. First $K$ draws are taken independently from the standard normal and collected in the $K\times 1$ vector $\mbf{z}$. A multivariate draw from $N(\bs{\mu},\bs{\Sigma})$ is then given by $\bs{\mu}+\mbf{L}\mbf{z}$.

\section{Application}

\subsection{English Longitudinal Study of Ageing (ELSA)}\label{ELSAdata}
To illustrate the methodology, longitudinal data are analysed from the English Longitudinal Study of Ageing (ELSA). The ELSA baseline (1998-2001) is a representative sample of the English population aged 50 and older.
Data from ELSA can be obtained via the  Economic and Social Data Service ({\tt www.esds.ac.uk}). There are 11932 individuals in the ELSA baseline.


Of interest for the current analysis is the change of cognitive function in older population.
For the current analysis, a random sample of size $N=1000$ is taken from ELSA. Of these 1000 individuals, 205 died during the follow-up with age at death available.
Because ELSA data are publicly available, measures have been taken by the data provider to prevent identification of the individuals. One of those measures is the censoring of ages above 90 years.  In the sampling of the subset of $N=1000$, individuals who were 90 years or older at baseline were ignored. 
The sample has  544  women and 456 men.

Highest educational qualification is dichotomised for the current analysis according to years of formal education:  fewer than ten versus ten or more. There are 558  individuals with fewer than ten years of education.

\subsection{A five-state model for remembering words}\label{sec4:ELSAapp}
\begin{figure}[t]
\includegraphics[width=290pt]{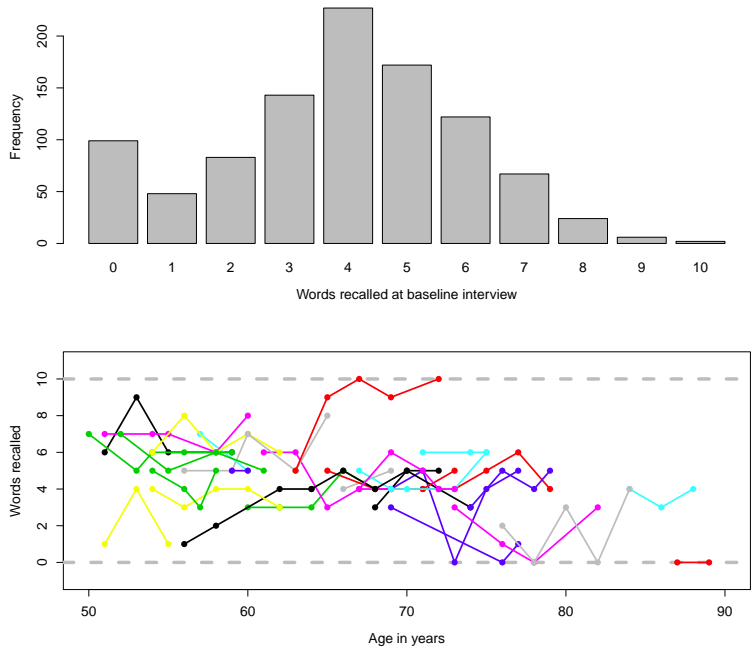}
\caption{Number of words remembered at the ELSA baseline (top graph), and follow-up trajectories  for a random subset of 30 individuals (bottom graph) 
\label{fig4:ELSAdata}}
\end{figure}

This application focuses  on the number of words remembered in a delayed recall from a list of ten.
 The score on this test is equal to the number of words remembered, i.e., score $\in\{0, 1, 2, \cdots, 10\}$.  The top graph in Figure \ref{fig4:ELSAdata} provides information on the number of words remembered at baseline. Most people remember 4 or 5 words, and the data show that remembering 9 or 10 words is exceptional. The bottom graph in  Figure \ref{fig4:ELSAdata} depicts the change of number of words remembered over time for a random subset of 30 individuals. The 30 trajectories illustrate that the delayed recall is a noisy process. Nevertheless, already in these trajectories  there is some evidence of a decline in cognitive function as people get older. The statistical modelling in this section aims to explore the effect of age and gender on cognitive change over time when controlling for education.

Four living states are defined by the number of words an individual can remember: state 1, 2, 3, and 4, for the number of words $\{7,8,9,10\}$,  $\{6,5\}$ $\{4,3,2\}$, and $\{1,0\}$, respectively. An additional state 5 is defined as the dead state, see also Figure \ref{figAvdH:ELSAdiagram}.

The interval-censored multi-state process is summarised by the frequencies in Table \ref{tab4:ELSAstatetable}. Note that the sum of the transitions into the dead state is equal to the number of deaths in the sample, i.e., 205.
Table \ref{tab4:ELSAstatetable} also shows that the process is mainly progressive in the sense that the main trend over time is towards the higher states.

\begin{table}
\caption{State table for the ELSA data: number of times each pair of states was observed at successive observation times. The four living states are defined by number of words remembered}
\setlength{\tabcolsep}{0.05 cm}
\centering
\begin{tabular*}{\columnwidth}{@{}r@{\extracolsep{\fill}}r@{\extracolsep{\fill}}r@{\extracolsep{\fill}}r@{\extracolsep{\fill}}r@{\extracolsep{\fill}}r@{\extracolsep{\fill}}l@{\extracolsep{\fill}}c@{\extracolsep{\fill}}c@{\extracolsep{\fill}}c@{}}
  \hline
  & \multicolumn{1}{l}{\textit{To}} & & & \\
\multicolumn{1}{l}{\textit{From}} &   \multicolumn{1}{l}{10-7 words}  &  \multicolumn{1}{l}{6-5 words}  &  \multicolumn{1}{l}{4-2 words}  & \multicolumn{1}{l}{1-0 words}  & \multicolumn{1}{l}{Dead} \\\hline
  10-7 words &  164& 150 & 49&   12 & 8\\
 6-5 words  &  156 &440& 303&  48&  40\\
  4-2 words   & 52 &336& 616& 151&  85\\
  1-0 words & 11 & 35& 114& 149 & 72\\\hline
\end{tabular*}
\label{tab4:ELSAstatetable}
\end{table}

In what follows, model estimation is undertaken by using the scoring algorithm.
Let ${\bs{\theta}}^\top=(\theta_1,...,\theta_q)$ be the vector with model parameters, where $q$ depends on the chosen model. The convergence criterion for the algorithm is to stop at iteration $v+1$ when $\sum_{k=1}^q |\theta_k^{(v)}-\theta_k^{(v+1)}| < 10^{-6}$.

Model selection is bottom-up starting with the time-homogeneous exponential hazard model given by
\begin{eqnarray}
q_{rs}(t)=\exp\big(\beta_{rs.0}\big),
\label{eq:ELSAmodel0}
\end{eqnarray}
for the transitions $r\rightarrow s$ depicted in Figure \ref{figAvdH:ELSAdiagram}. This intercept-only model with 10 parameters has AIC = 8109.5. Convergence of the scoring algorithm was reached after 14 iterations, using starting values $\beta_{rs.0}=-3$ for all the parameters.

For the process at hand, age is the most suitable time scale. Age in the ELSA data is transformed by subtracting 49 years. This results in 1 being the minimal age in the sample.

Even though the sample size is not small, Table \ref{tab4:ELSAstatetable} shows that mortality information is limited because only about 20\% of the individuals end up in the dead state during follow-up.
In what follows, model (\ref{eq:ELSAmodel0}) is extended by adding parameters with parameter equality constraints. The first extension is a Gompertz model given by
\begin{eqnarray}
q_{rs}(t)=\exp\big(\beta_{rs.0}+\xi_{rs}t\big),
\label{eq4:ELSAmodel2}
\end{eqnarray}
where $\xi_{21}=\xi_{32}=\xi_{43}=\xi_B$ and $\xi_{15}=\xi_{25}=\xi_{35}=\xi_{45}=\xi_D$. That is, the effect of time is the same for all backwards transitions and for transitions into the dead state. In the estimation, the grid for the piecewise-constant approximation is defined by individually observed follow-up times in the data.
This model has 15 parameters, and needs 16 scoring iterations when using starting values $\beta_{rs.0}=-3$ and $\xi_{rs}=0$ for all the relevant $r,s$-combinations. The model has AIC =  7780.5.

Subsequently, covariate information is added for the transitions of interest, i.e., those transitions that represent a decline in cognitive function. For this, model (\ref{eq4:ELSAmodel2}) is extended to
\begin{eqnarray}
q_{rs}(t)=\exp\big(\beta_{rs.0}+\xi_{rs}t+\beta_{rs.1}sex+\beta_{rs.2}education\big),
\label{eq4:ELSAmodel3}
\end{eqnarray}
where $sex$ is 0/1 for women/men, and $education$ is 0/1 for fewer than ten years/ten years of more of education. For the transitions into the dead state, the constraints on the coefficients for $sex$ are $\beta_{15.1}=\beta_{25.1}=\beta_{35.1}=\beta_{45.1}$, and for $education$ are $\beta_{r5.2}=0$ for $r=1,2,3,4$.
This model has 22 parameters, needs 16 iterations, and has AIC = 7680.3.

It is worthwhile to investigate alternative time-dependent models. First, in model (\ref{eq4:ELSAmodel3}), the Gompertz baseline models for the transitions into the dead state are replaced by Weibull models.
Starting values for the transitions into the dead state are $\beta_{r5.0}=-10$, $\tau_{15}=\exp(0.5)$, and for the remaining parameters the values are as given above.
This yields AIC = 7688.7 after 20 iterations.

Next, all baseline hazards definitions in model (\ref{eq4:ELSAmodel3}) are replaced by Weibull models, which results in AIC = 7729.5 after 28 iterations. Alternatively, model (\ref{eq4:ELSAmodel3}) is defined with Gompertz baseline models for the transitions into the dead state and Weibull models for progression through the living states. This yields AIC = 7719.7 after 25 iterations.

\begin{table}[t!]
\caption{Comparison between models for the ELSA data with $N=1000$, where -2LL stands for -2 times the (penalised) loglikelihood function evaluated at its maximum}
\centering
\begin{tabular*}{\columnwidth}{@{}l@{\extracolsep{\fill}}l@{\extracolsep{\fill}}c@{\extracolsep{\fill}}r@{\extracolsep{\fill}}r@{\extracolsep{\fill}}r@{\extracolsep{\fill}}l@{\extracolsep{\fill}}c@{\extracolsep{\fill}}c@{\extracolsep{\fill}}c@{}}
  \hline
Model & \multicolumn{1}{l}{Baseline hazards} & \multicolumn{1}{l}{\#Parameters} & \multicolumn{1}{l}{-2LL} & \multicolumn{1}{l}{AIC}\\ \hline
  Intercept-only & Exponential&  10& 8089.5& 8109.5\\
  $t$                 & Gompertz &  15& 7750.5& 7780.5\\
  $t$, $sex$, $education$ & Gompertz &  22& 7636.3& 7680.3\\
  $t$, $sex$, $education$ & Gompertz for living  &  22& 7644.7	& 7688.7\\
                           & \multicolumn{2}{l}{\ \ and Weibull for death} & & \\
  $t$, $sex$, $education$ & Weibull &  22& 7685.5& 7729.5\\
  $t$, $sex$, $education$ &  Weibull for living &  22& 7675.7& 7719.7\\
                           & \multicolumn{2}{l}{\ \ and Gompertz for death} &  \\
 $t$, $sex$, $education$ & $P$-splines I  &  38& 7626.0 & 7678.2 \\
                           & \multicolumn{2}{l}{\ \ for $2\rightarrow3$ and $3\rightarrow4$} &  \\
 $t$, $sex$, $education$ & $P$-splines II for $3\rightarrow4$  & 30& 7630.9 & 7678.2\\  \hline
\end{tabular*}
\label{tab4:AICs}
\end{table}

Semi-parametric models with $P$-splines can be used to model non-linear functional forms and to check shapes specified by parametric models. To illustrate this, the Gompertz  hazard for transition $2\rightarrow 3$ and $3\rightarrow 4$ in model (\ref{eq4:ELSAmodel3}) are replaced by

\begin{equation}\label{eq:semp1}
\begin{split}
q_{23}(t) &= \exp\left(\sum_{k=1}^{K}\alpha_{23.k}B_k +\beta_{23.1}sex+\beta_{23.2}education \right) \\
q_{34}(t) &=\exp\left(\sum_{k=1}^{K}\alpha_{34.k}B_k +\beta_{34.1}sex+\beta_{34.2}education \right).
\end{split}
\end{equation}

For this model, the number of $P$-splines bases for both hazard functions is $K=10$ and the vector of smoothing parameters is  $\mbox{\boldmath$\lambda$}^\top = (\lambda_1, \lambda_2)$. The initial grid is given by all pairs of combinations of $\log_{10}\lambda_1 = \{-3,-2,-1,0,1,2,3\}$ and $ \log_{10} \lambda_2 = \{-3,-2,-1,0,1,2,3\}$. A possible graphical representation of the AIC results is to plot its values when one smoothing parameter is fixed. Figure \ref{fig:results1}(c) illustrates the resulting AIC for different values of $\lambda_2$ with fixed $\lambda_1 = 10^{-3}$. The value which minimises the AIC is $\lambda_2 = 10$. It happens for all values of $\lambda_1$. The search for the optimal values of $\lambda_1$ is less straightforward as $\lambda_1 \rightarrow \infty$. Figure \ref{fig:results1}(a) shows the AIC  for several values of $\lambda_1$ with fixed $\lambda_2 = 10$. The AIC decreases quickly for small values of $\lambda_2$; however, it gets constant for large values. This result indicates that the functional form of the hazard for transitions $2 \rightarrow 3$ is $\log$-linear. Because both AIC and parameter estimates do not change much for sufficiently large values of $\lambda_1$, it is possible to set $\lambda_1 = 10^7$. In this case, the best model ($P$-splines I) according to the AIC is obtained with smoothing parameter  $\widehat{\mbox{\boldmath$\lambda$} }^\top = (10^7, 10)$. This model has 30 parameters and 26.1 degrees of freedom. 

The fitted hazards for transition $2\rightarrow 3$ for the Gompertz (\ref{eq4:ELSAmodel3}) and $P$-splines I (\ref{eq:semp1}), for men with ten or more years of education are illustrated in Figure \ref{fig:results1}(b). The functional forms of both models are very similar for this transition; however, the functional forms for transition $3\rightarrow 4$ are quite different, as indicated in Figure \ref{fig:results1}(d). Model (\ref{eq:semp1}) has AIC = 7678.2 indicating that it performs better than the Gompertz model with AIC = 7680.3.

The functional form of hazard for transition $2\rightarrow 3$ in model (\ref{eq:semp1}) indicates that a Gompertz specification can be reasonable for this transition. Therefore, in model (\ref{eq4:ELSAmodel3}), only the hazard for transition $3\rightarrow 4$ is specified with $P$-splines:
\begin{eqnarray}
q_{34}(t) &=& \exp\left(\sum_{k=1}^{K}\alpha_{34.k}B_k +\beta_{34.1}sex+\beta_{34.2}education \right).
\label{eq:semp2}
\end{eqnarray}

The number of $P$-splines bases is $K = 10$ and the grid search is made on the values $\log_{10} \lambda = \{-3, -1, 0, 1, 3\}$. The resulting AIC values are illustrated in Figure \ref{fig:results1}(e). The minimum AIC with value 7678.2 is obtained at $\lambda = 10$.  That is the same AIC value as for model (\ref{eq:semp1}); however, the degrees of freedom is slightly smaller $df=23.65$. As model (\ref{eq:semp2}) ($P$-splines II) is easier to estimate if compared to model (\ref{eq:semp1}), it is considered the best model among all illustrated in this paper. Table \ref{tab4:AICs} summarises the comparison of the investigated models. Figure \ref{fig:results1}(f) illustrates the fitted hazard for transition $3\rightarrow 4$ in model (\ref{eq:semp2}) for men with ten or more years of education. As expected, there is an increase of risk of progression to a decline of cognitive function over the years.

\begin{figure}[t]
\center
\includegraphics[width=11cm]{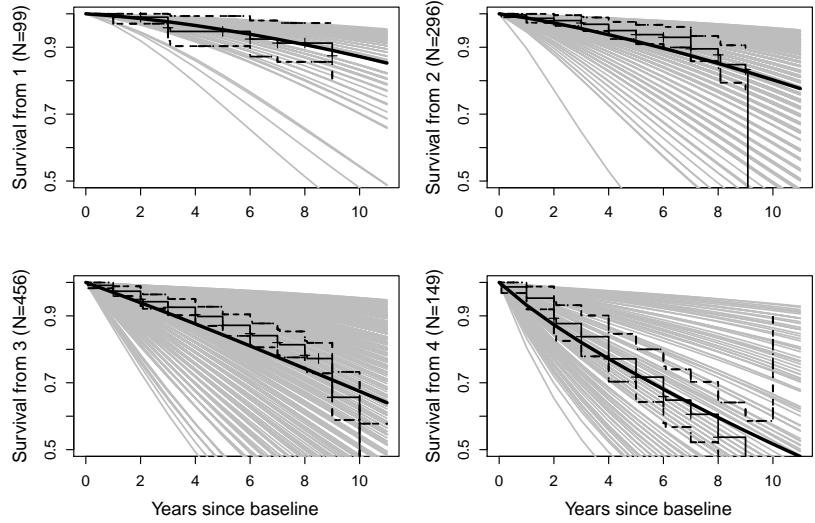}
\caption{Comparison of model-based survival from states 1, 2, 3, and 4 with Kaplan-Meier curves. Model-based survival: grey lines for individuals, smooth black line for the mean of the individual survival curves. Kaplan-Meier in black lines with 95\% confidence bands. Frequencies for baseline state along vertical axes \label{fig:ELSAGOF}}

\end{figure}

\begin{figure}[!t]
\center
\includegraphics[width=4.8in]{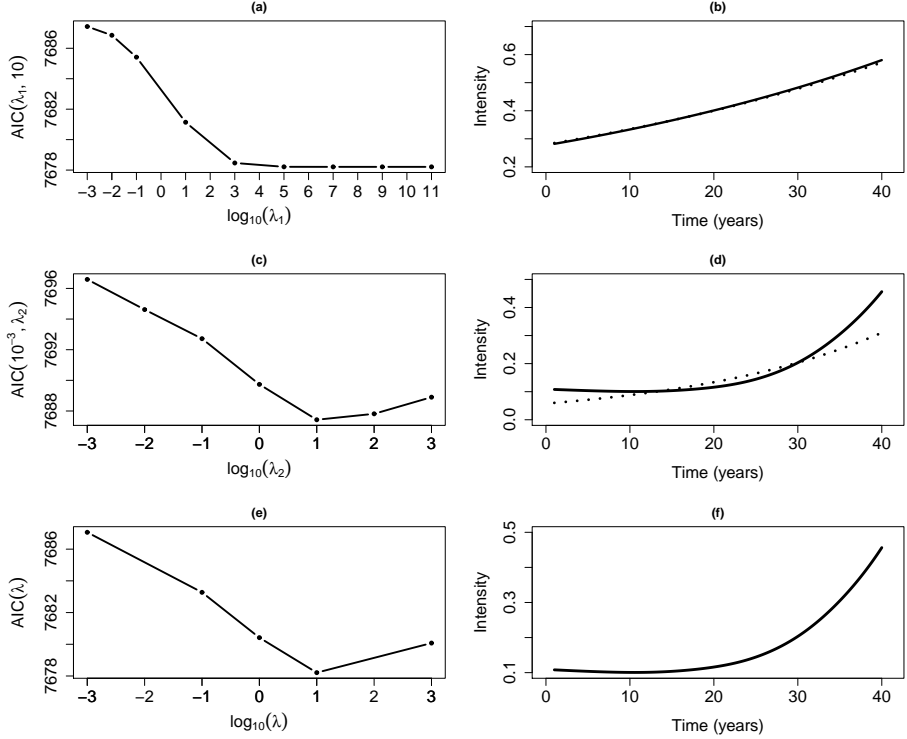}
\caption{AIC results and fitted hazard transitions for men with ten or more year of education. In (a) and (c), the AIC results for fixed $\lambda_2 = 10$ and fixed $\lambda_1 = 10^{-3}$, respectively. In (b) and (d), the estimated hazards for  $2\rightarrow 3$ and $3\rightarrow 4$, respectively. Solid line for $P$-splines I and dotted line for Gompertz. In (e) and (f), the AIC results for model $P$-splines II and fitted hazard for $3\rightarrow 4$, respectively 
\label{fig:results1}}
\end{figure}

Model validation is hampered by the interval censoring of the transitions between the living states. But given that death times are available, it make sense to compare survival as estimated by the model with Kaplan-Meier curves  \cite{gentleman1994}. Of course, this will only check part of the fitted model. Figure \ref{fig:ELSAGOF} depicts baseline-specific survival as estimated by the model and as described by the Kaplan-Meier curves. For survival given baseline state 3, there is some discrepancy between model-based mean survival and the Kaplan-Meier curve, but overall the fit is reasonably good. Although this is not a proper goodness-of-fit test, the comparison shows that the model is able to capture the attrition due to death during the follow-up.

Table \ref{tab4:ELSA} illustrates the parametric component estimates for the $P$-splines II model. Most of the point estimates are according to expectation. For example, the effect of getting older is associated with decline of cognitive function, i.e.,  $\widehat{\xi}_{12}$, and with a decreasing hazard of remembering more words, i.e., $\widehat{\xi}_{B} <0$. For transitions $1\rightarrow 2$, $2\rightarrow 3$, and $3\rightarrow 4$ more years of education is associated with a lower risk of moving.

\begin{table}[t!]
\caption{Results for sex, education and time for the five-state $P$-splines II model for the ELSA data. Estimated standard errors in parentheses}
\setlength{\tabcolsep}{0.25 cm}
\begin{center}
\begin{tabular*}{\columnwidth}{@{}l@{\extracolsep{\fill}}r@{\extracolsep{\fill}}l@{\extracolsep{\fill}}r@{\extracolsep{\fill}}l@{\extracolsep{\fill}}r@{\extracolsep{\fill}}l@{\extracolsep{\fill}}c@{\extracolsep{\fill}}c@{\extracolsep{\fill}}c@{}}
  \hline
\multicolumn{1}{l}{$sex$} &   &\multicolumn{1}{l}{$education$} &  & \multicolumn{1}{l}{$t$} & \\ \hline
 $\beta_{12.1}$ &0.552  \ (0.138) &$\beta_{12.2}$ &-0.281 \ (0.146)&  $\xi_{12}$ &0.030 \ (0.010)  \\
 $\beta_{23.1}$ &0.178  \ (0.101) &$\beta_{23.2}$ &-0.836 \ (0.103) & $\xi_{B}$ &-0.031 \ (0.006)  \\
 $\beta_{34.1}$ &0.141  \ (0.145) & $\beta_{34.2}$ &-0.445 \ (0.160) & $\xi_{D}$ &0.042  \ (0.009)\\
 $\beta_{D}$ &0.477  \ (0.151) & & & \\
\hline
\end{tabular*}
\end{center}
\label{tab4:ELSA}
\end{table}

\subsection{Predicting cognitive function }
Although parameters for the transition intensities help to understand the estimated model, interpretation is more straightforward when transition probabilities are considered.
Firstly, consider a short time interval for which we assume that the intensities are constant. For men aged 60 with ten or more years of education, the two-year transition probabilities are estimated at
\begin{eqnarray*}
\widehat{\mbf{P}}\left(t_1=11,t_2=13\Big|\begin{tabular}{l}$sex=1,$\\$education=1\!\!$\end{tabular}\!\!\right)=
{\small
\left(
\begin{array}{lllll}
  0.330 &0.488 &0.154& 0.010& 0.018\\
  0.171 &0.531 &0.253 &0.023 &0.022\\
  0.083& 0.391 &0.429 &0.066& 0.031\\
  0.034 &0.219& 0.410 &0.291 &0.046\\
 0& 0& 0& 0& 1
\end{array}
\right),
}
\end{eqnarray*}
where $t$ denotes age transformed by subtracting 49 years. The diagonal entries in this matrix dominate. 
But there are some large off-diagonal entries as well. For example, if a man aged 60 is in state 3, then he has a 39\% chance of being in state 2 two years later. This high chance is an illustration of the noisiness of the process under investigation: it is quite likely that a 60 year old man moves between states 2 and 3 within the next two years.

\begin{figure}[t]
\center
\includegraphics[width=345pt]{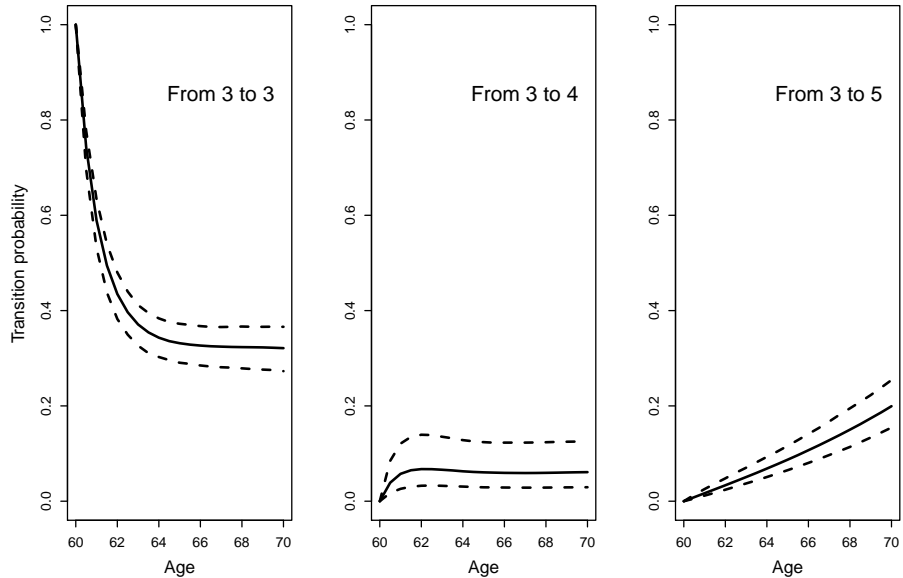}
\caption{For the $P$-splines II model, estimated ten-year transition probabilities for men aged 60 with ten or more years of education, and in state 3 at baseline. Solid line for transition probabilities (with $B=1000$) and dashed lines for 95\% confidence bands \label{fig:ELSAProbs}}

\end{figure}

Next we illustrate the estimation of standard errors and 95\% confidence intervals for transition probabilities.
Using simulation with $B=1000$, we obtain
\begin{eqnarray*}
\widehat{\mbf{P}}\left(t_1=11,t_2=13\Big|\begin{tabular}{l}$sex=1,$\\$education=1\!\!$\end{tabular}\!\!\right)=
{\small
\left(
\begin{array}{lllll}
  0.330 & 0.484& 0.153 &0.011& 0.021\\
 0.170 &0.529 &0.253 &0.024& 0.024\\
0.083& 0.390 &0.427 &0.069 &0.032\\
 0.034 &0.219 &0.407& 0.292& 0.048\\
 0 &0 &0 &0 &1
\end{array}
\right),
}
\end{eqnarray*}
with estimated standard errors
\[
{\small
\left(
\begin{array}{ccccc}
 0.038& 0.029& 0.016& 0.004& 0.011\\
 0.013 &0.021 &0.019 &0.009 &0.006\\
 0.007 &0.019 &0.024 &0.024 &0.006\\
 0.004 &0.019 &0.024 &0.037 &0.012
\end{array}
\right).
}
\]
The 95\% confidence intervals for the first row are given by
\[
{\small (0.263, 0.408), \   (0.425,0.540), \  (0.120,0.185), \  (0.004,0.021),}  \mbox{\ and\ }
{\small (0.011,0.049)}.
\]
Next, ten-year transition probabilities are estimated for men aged 60 with ten or more years of education. The grid is defined by $h=1/2$ years. The estimation is shown in Figure \ref{fig:ELSAProbs}. 

Figure \ref{fig:ELSAProbs} concurs with the expectations. For example, given the progressive trend of the process, it is to be expected that probability of being in state 3 decreases over time, as moving to states 4, and 5 becomes more likely due to increased age.

\section{Discussion}
\label{s:discuss}
Specification and estimation of continuous-time multi-state survival models are presented and shown to be a flexible framework for statistical modelling of time-dependency processes. By defining transition-specific parametric and semi-parametric hazard models, a wide range of multi-state processes can be investigated. Penalised maximum likelihood estimation is undertaken by a scoring algorithm using a piecewise-constant approximation to time-dependent hazards. The Akaike information criterion is used to select the optimal value for the smoothing parameters.

The Markov process formulation to semi-parametric multi-state models extends the method described in Joly and Commenges \cite{joly1999penalized} and Joly \textit{et al.} \cite{joly2002penalized}. This is an important methodology to medical studies as backwards transitions occur naturally in many applications \cite{abner2012, marioni2012}. Furthermore,  using the piecewise-constant approximation is an alternative to the method introduced by Titman \cite{titman2011} which handles the  time-dependency by using numerical solutions to the non-linear differential equations which are defined directly by the time-dependency of the Markov process. As stated by Titman, computation using the non-linear differential equations can become prohibitively slow when adding continuous covariates. This is not a problem when using the piecewise-constant approximation and the scoring algorithm.

The scoring algorithm is implemented in {\sf R} in such a way that it is easy to vary transition-specific choices for parametric and semi-parametric shapes. An example of such a model is explored in the application, where $P$-splines are used for transitions $2\rightarrow 3$ and $3\rightarrow 4$, and Gompertz hazards are defined for the other transitions. The eigenvalue decomposition in the algorithm is computed with the function {\tt eigen} in {\sf R}, which uses the LAPACK routine \cite{anderson1999}. $P$-spline bases are computed using the code in the appendix in Eilers and Marx \cite{eilers1996}.

If prediction of a time-dependent process beyond the time range in the data is of interest, hazard models with $P$-splines can be used to validate the parametric choices which underlie the prediction. This was illustrated in the application with the ELSA data in which age range is from 50 to 90 years.  If risk factors are the main focus of the research, $P$-splines can be used to capture non-parametric shapes of time-dependency.

The choice of the type of spline is not essential. $P$-splines were used in this paper, but any other spline function with a first-order derivative can be handled within the current framework. The same holds for parametric shapes other than the Gompertz and the Weibull. The specification and estimation of continuous-time survival model is very general and does not pose restrictions on the number of states, scale of covariates, or number of transitions.


\section*{Acknowledgements}

The ELSA data were made available through the UK Economic and Social Data Service. ELSA was developed by a team of researchers based at the National Centre for Social Research, University College London and the Institute for Fiscal Studies. The developers and funders of ELSA do not bear any responsibility for the analyses or interpretations presented here. This research was supported by CNPq - Brazil [249308/2013-4].
\vspace*{-8pt}


\bibliography{mybibfile}

\end{document}